**Title**: Flow reactor for preparation of lipid nanoparticles via temperature variations

**Authors**: I. Lesov, [1] D. Glushkova, [1] D. Cholakova, [1] M.T. Georgiev, [1] S. Tcholakova, [1] S. K. Smoukov, [2] N. Denkov[1*]

[1]*Department of Chemical and Pharmaceutical Engineering, Sofia University, 1 J. Bourchier Ave., 1164 Sofia, Bulgaria*

[2]*School of Engineering and Materials Science, Queen Mary University of London, London E1 4NS, U.K.*

*Corresponding author:

Prof. Nikolai Denkov

Department of Chemical and Pharmaceutical Engineering

Faculty of Chemistry and Pharmacy, Sofia University

1 James Bourchier Ave., 1164 Sofia

Bulgaria

Phone: (+359-2) 962 5310

Fax:    (+359-2) 962 5643

E-mail: ND@LCPE.UNI-SOFIA.BG

**Abstract**

Lipid nanoemulsions and nanosuspensions are used as flavor carriers and bubble stabilizers in soft drinks and foods, as well as delivery vehicles for lipophilic drugs in pharmaceutics. Common techniques for their formation are the high-pressure and ultrasonic homogenizers. These techniques dissipate most of the input energy, which results in excessive heating and generation of free radicals that might modify sensitive ingredients. Low energy methods are also used in some applications, but they have specific limitations restricting their universal use. In the current study, we propose an alternative approach - a flow reactor with a variable temperature, which utilizes the lipids' polymorphic transitions to induce spontaneous fragmentation of the lipid microparticles into nanoparticles. The reactor allows us to obtain emulsions or suspensions with particle diameters tunable between 20 and 800 nm when appropriate surfactants, temperature profiles, and flow rates are applied. The fragmentation is comparable to that in a high-pressure homogenizer at ca. 500 bars or higher, without creating emulsion overheating or cavitation typical for the conventional methods. The flow reactor can be scaled up to industrial applications using simple scaling rules.

**Highlights:**

A scalable flow reactor for the production of lipid nanoemulsions and nanosuspensions is described

The reactor uses temperature variations for spontaneous fragmentation of coarse lipid particles

Lipid nanoparticles with a diameter down to 20 nm were obtained

The size of the obtained particles is similar to that obtained in a high-pressure homogenizer

The main factors for process optimization and scaling are clarified

1. **Introduction**

The preparation of lipid nanosuspensions and nanoemulsions is essential for many foods [1] and pharmaceutical applications [2]. The lipid nanoemulsions are often applied in beverages [3-4] and foods [5], among which soft drinks and ice cream, where the fat particles serve as flavor carriers [6] and as structural units [7]. The lipid nanoparticles are also used and studied intensively as drug delivery vehicles for controlled drug release [2,8].

The commercial preparation of nanoemulsions and nanosuspensions can be achieved through a palette of different technologies [9], such as high energy dissipation devices that include high-pressure, ultrasonic [10-11], and rotor-stator homogenizers [12-14]. Low energy methods are also used in some specific applications [15], where the formation of the nanoparticles occurs via catastrophic changes in the emulsion stability with related phase inversion (e.g., from water-in-oil into oil-in-water) driven by temperature variations or composition changes.

A common technique in food and pharma industries is high-pressure homogenization (HPH) [16] which produces submicrometer droplets from low- and medium-viscosity oils. HPH requires heavy maintenance, and often results in undesired phenomena that include cavitation and overheating at elevated pressures that might deteriorate sensitive components such as flavors, vitamins, enzymes, and other proteins [17]. Similar concerns apply to ultrasonic technologies, where additional problems arise from the more common batch processing than the preferred continuous flow regime [18-19].

Rotor-stator devices have been also used for the formation of nanoemulsions during the last decade [12-14]. The formation of nanoemulsions via rotor-stator homogenization requires a viscosity enhancement to realize "viscous turbulent regime" [14] that allows effective breakage of the droplets. As a result of the viscosity increase and high shear rates, intensive energy dissipation is observed that can lead to undesired emulsion overheating [14]. On the other hand, the low-energy (temperature or composition) phase inversion techniques require fine-tuning for different systems that are best done in batches [20]. Another issue accompanying this technique is the variation of the commercial oil grades and purity [21-22], which often obstructs the method's reproducibility across different manufacturing sites. Additional problems arise from the often-needed solvent enhancing additives that are subject to different regulations in the various countries and sometimes face exceptionally costly and prolonged health assessment procedures.

Here we present an alternative process for producing lipid nanoparticles and nanodroplets based on the polymorphic phase transitions inherent to the lipid compounds [23-26]. Recently Cholakova et al. [25-26] showed that triglyceride emulsions, subject to freeze-thaw cycles of the dispersed particles, can yield nanoemulsions via spontaneous bursting of the solid triglyceride particles. In this approach, we needed no mixer to induce particle fragmentation because the initial coarse particles burst into millions of nanoparticles in a single cooling-heating cycle. The nanoemulsions formation depended exclusively on selecting appropriate surfactants and required a rapid cooling of the coarse lipid particles below their freezing point, followed by slow heating, as illustrated in Fig. 1. The versatility of the method was demonstrated with a variety of different lipids, including natural oils (cocoa butter, coconut oil, and palm kernel oil) and lipids of high interest to the pharmaceutical industry (such as Precirol ATO5 and Gelucire 43/01) while using surfactants that are widely used in the food, cosmetic and pharmaceutical industries [26].

In the current study, we realize this process in a pilot-scale flow reactor that can generate up to 30-60 milliliters of nanoemulsion per hour and can be up-scaled to an industrial level by using simple scaling rules based on the optimized temperature gradients and residence times.

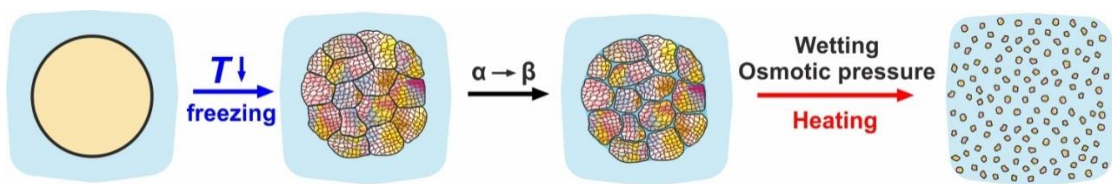

**Figure 1**. Illustrative scheme of the lipid particle fragmentation via polymorphic phase transitions in a cooling-heating cycle (adapted from Ref. [25]). The lipid drops freeze upon rapid cooling, and upon storage or subsequent slow heating, polymorphic phase transitions cause the formation of nanovoids in the lipid particles, which are flooded by the surfactant solution. The surfactant wets the nanochannels, causing a fragmentation of the solid particle into millions of smaller particles. As a result, nanodispersions are formed.

## 2. Methods and materials

*Coarse emulsions* with mean volume-surface drop diameter, $d_{32}$, falling in the range between ca. 5 and 10 μm, were prepared via simple rotor-stator homogenization (Ultra Turrax T25 Digital, IKA). We mixed the water and oil phases with both surfactants pre-added in the water phase at temperature 5 to 10 °C above the oil melting point at 12 000 ± 1 000 rpm for 5 min and stored for up to 1 week at a temperature above the oil melting point, before using them for nanoemulsion formation.

**The continuous flow reactor** resembles some lab-on-a-chip devices that allow precise temperature control and tracking in micrometer channels [27]. However, the dimensions of the channels in our device are much larger to increase its production capacity. The device is illustrated in Figure 2A-B and consists of a rectangular serpentine channel with dimensions 1350 mm in length, 3.2 mm in height, and 3.4 mm in width (a total inner volume of 14.7 mL). 3M Adhesive Transfer Tape 467MP was glued to a 3 mm PMMA plastic sheet, and channels were cut into the intermediate PMMA channel via laser to ensure low surface roughness. The channel plate was then glued to two other 3 mm PMMA plates – one above and one below. The lower plate had 1 mm deep rectangular carvings on its bottom that were filled with thermal paste, and where 6 Peltier elements (TES1-12703, Hebei I.T.) were set as illustrated in Figure 2, 15 mm apart from each other. The temperature of the Peltier elements was controlled via a commercial controller, TEC-1123-SV (Meersteter Engineering, Switzerland), equipped with TEC 3.0 software and four external Pt1000 sensors, located between the Peltier elements and the lower PMMA plate below each corner Peltier. The Peltier elements were cooled via alumina heatsinks, connected to a Julabo Corio CD circulator. The emulsion cycling through the channels was ensured via syringe pump, NE-1000 (New Era Pump Systems Inc., USA). The injected coarse emulsion and the final nanoemulsion were kept at ambient temperature, $26 \pm 3°C$

Before beginning each experiment, the reactor was filled with deionized water. Then, its temperature zones were adjusted to the desired ranges and left to equilibrate for 10 minutes. Next, deionized water was injected for 10 min at the pumping rate of the experiment, ensuring that the temperature gradients in the cell would not vary significantly during the emulsion pumping. Finally, we connected a syringe with 25 mL of coarse emulsion and circulated it through the cell by gradually displacing the water in the cell. As the reactor allowed for nearly plug flow conditions, only the emulsion front was diluted by the water initially present in the channel. Thus, we discarded the first 1.5 mL of the emulsion before collecting the rest of the sample. Once the syringe with the coarse emulsion was fully injected, we connected another syringe with air to pump out the remaining emulsion. As soon as we collected at least 15 mL of the emulsion, we took a 1 mL sample for particle size characterization via optical microscopy and DLS. The procedure was repeated for up to 4 cycles.

After use, the cell was cleaned via soaking in commercial detergent at an elevated temperature and then flushing it several times. Afterward, it was thoroughly rinsed with deionized water. Organic solutions, such as ethanol, were avoided due to the formation of cracks in the PMMA sheets.

**I.R. thermal imaging** was used to optimize the temperature gradients and to measure the temperature within the channels [28]. The I.R. camera used (Flir One Pro Gen 3 for Android with USB-C, Flir systems) has a spatial resolution of 640×480 pixels (≈ 0.5 mm/pixel in the most images taken) and a thermal resolution of 0.1 °C. Its reported accuracy is 3-5% from the camera-to-sample temperature difference, which corresponds to ± 2 °C at the temperature extremes that we achieved in our experiments. To reduce the error to 0.5 °C or less, we performed two-point temperature calibration for each thermal image, which we used to correct the profiles via the Flir Tools software. Then, we calibrated the reactor's surface temperature against a Pt1000 sensor inserted in the internal fluid during constant flow. Both procedures are described in detail in *S.I. section S1*. After extracting the complete profile of the fluid temperature in the cell, the measured oscillations were mathematically smoothed using a local linear regression model with a sampling proportion of 0.05 or 0.10. An illustrative result for the temperature achieved against the temperature set and the local smoothening is given in Figure 2C. However, the residence times in each temperature range were calculated from the original non-smoothed data to avoid errors near the fragmentation and melting temperature transitions.

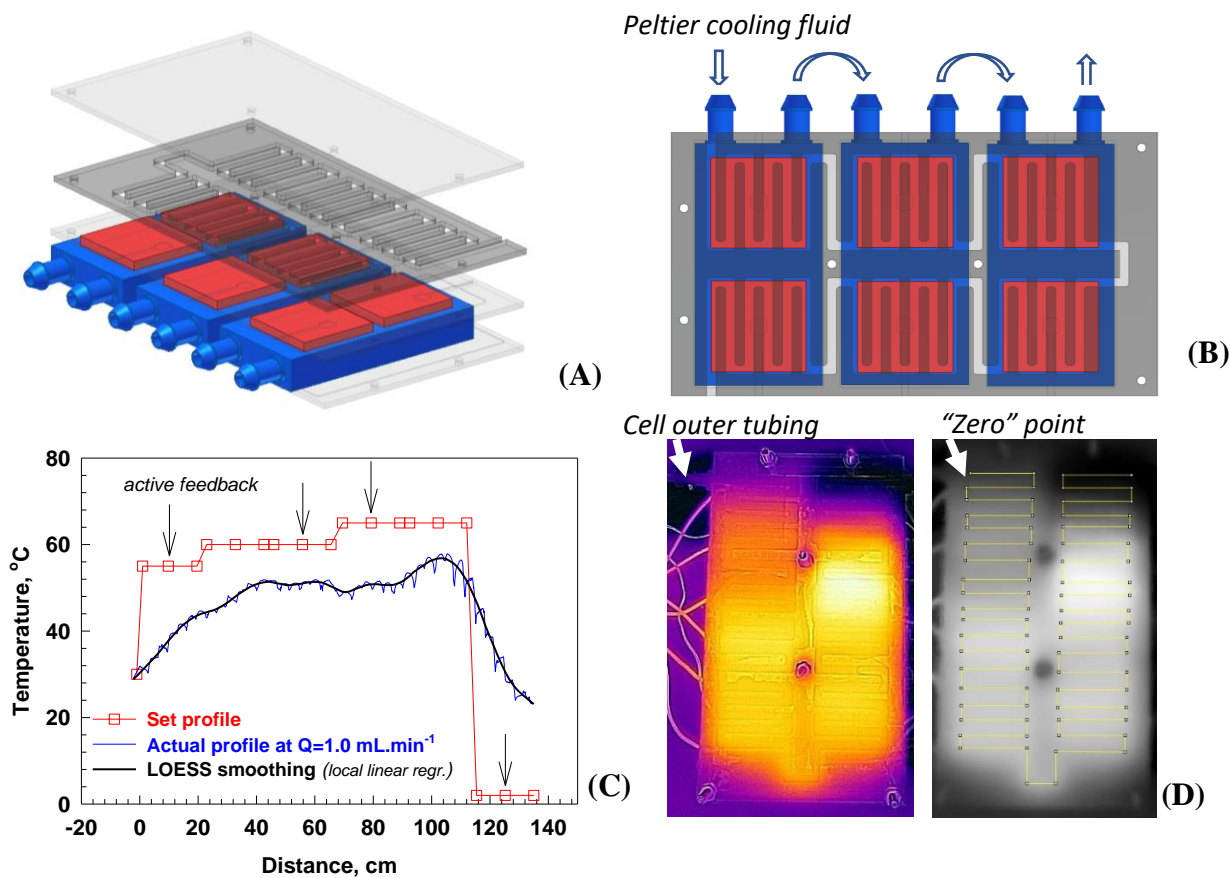

**Figure 2**. Schematics of the reactor primary cell: **(A)** Exploded view - the reactor consists of a 3-layer PMMA plastic with meandering, laser-cut channels in the mid-plate. There are 6 Peltier elements below the lower plate marked in red; 4 are connected to a Pt1000 sensor below the plastics, ensuring active feedback. The Peltier elements are controlled via TEC-1123-SV boards (Meersteter Engineering, Switzerland), and their lower side temperature is set constant via alumina heatsinks connected to a chiller. **(B)** Reactor top view – serpentine flow channels with average dimensions 1350×3.2×3.4 *(l×h×w)* mm$^3$ go over the Peltier elements. **(C)** Temperature profile set in the 6 Peltier elements (software input sequence 55-60-60-65-65-2 °C) and actual temperature profile, measured at a flow rate of 1 mL.min$^{-1}$. The profile is extracted using a calibrated infrared camera and is smoothed via local linear regression, as described in the text. **(D)** The blended I.R. - optical image is shown in a rainbow palette to illustrate channels and temperature, whereas the black and white thermograph shows the whole temperature extraction path. The heat from the channels is conducted via the outer tubing, which changes the emulsion temperature above the ambient one at the zero point.

***High-pressure homogenization*** was made for comparison with the same emulsions using lab-scale homogenizer PandaPLUS 2000, GEA. The coarse emulsions were homogenized at pressures from 300 to 750 bars, passing two times at a given pressure, and samples were taken for size measurements. To avoid clogging of the homogenizer in the case of lipid solidification, we preheated the HPH and the emulsions about 10-15°C above the oil melting point.

***Drop sizes*** in the studied emulsions were characterized via two complementary methods – optical microscopy (O.M.) and dynamic light scattering (DLS).

*A transmitted light optical microscope* (Axioplan, Zeiss) was used for observation of non-diluted emulsions after the rotor-stator homogenization (coarse emulsions) and after passing through the reactor (to check for micrometer-sized particles). The emulsions were placed between two glass slides and were observed with a 50× LD Epiplan objective. At least 1000 drops were measured for a given sample, and their Sauter diameter was calculated.

*Light scattering* (Nanosizer Nano Z.S., Malvern) equipped with 633 nm laser and working at 173° backscattering mode and software version 7.13 was used for drop size evaluation. About 1 mL of deionized water was placed in a glass cuvette, and 0.05 g of the emulsion was diluted in it. Then, the sample was sealed and gently turned upside down 2-3 times to homogenize it. An automated selection of optics and attenuators was allowed. The data were interpreted via multimodal analysis and averaged from at least three measurements from each sample and 2-4 independent sample repetitions to check for reproducibility. The mean diameter varied within 30% between independent samples and across different runs for one sample. The measurements were performed at 25 °C.

In part of the experiments, the DLS was also used to optimize the temperature intervals, where fragmentation occurs for the suspended nanoparticles. Nanoemulsions, prepared after two cycles of fragmentation, were subject to size measurements at different temperature intervals, as explained below.

***Differential scanning calorimetry*** (DSC 250, T.A. Instruments) was used to characterize the oil melting regions needed to set the reactor's temperature gradients. The bulk oil samples were melted, measured in Tzero pans, and then subjected to a heating-cooling-heating sequence at a constant rate of 5 °C.min$^{-1}$.

***SAXS/WAXS*** measurements were performed using the Xeuss 3.0 (Xenocs, France) apparatus using a Cu radiation source, $\lambda \approx 1.5419$ Å. The signal was detected with a EIGER2 R 4M detector at a sample-to-detector distance of 286 mm. Exposure time for a single scattering profile was 200 s. The temperature of the sample was controlled using Linkam high-temperature stage (MFS350, Linkam, UK).

*Materials.* Edible or pharmaceutically allowed oils and nonionic surfactants were selected for the current process upscaling. The oils used in this study were Precirol ATO 5 (Gattefosse, Germany), Coconut oil (Smart Organic, Bulgaria), Palm kernel oil (BiOrigins, UK) and trimyristin (≥ 95 %, TCI, Japan). The surfactants polyoxyethylene (20) sorbitan monolaurate (Tween 20, Sigma-Aldrich), 1-oleoyl-rac-glycerol 40 %, Sigma-Aldrich; polyoxyethylene (20) stearyl ether (Brij S20, Sigma-Aldrich) and polyoxyethylene (4) lauryl ether (Brij 30, Sigma-Aldrich) were used for emulsion stabilization. The 1-oleoyl-rac-glycerol 40 % was recrystallized twice in ethanol to yield 90 % purity of monoolein as in Ref. [26], while the rest of the chemicals were used as received. The aqueous solutions were prepared with deionized water with resistivity > 18 MΩ.cm, purified by Elix 5 module (Millipore) and analytical grade NaCl from Sigma-Aldrich were used. The materials are described in more detail in Ref. [26].

### 3. Experimental results

First, we describe the results obtained with initial coarse emulsions of 1 wt% Precirol as the lipid phase and the following surfactants: 1.5 wt% Brij S20 and 0.5 wt% Brij 30 added to the phase. We subjected these emulsions to different numbers of passes through the reactor, different flow rates, and different temperature gradients to characterize the fragmentation efficiency in the flowing emulsions and determine the sizes of the obtained drops, depending on the temperature gradients set in the reactor.

Figure 3A illustrates the smoothed temperature profiles in the channel, depending on the flow rate in the reactor at the same temperature sequence: 55-60-60-65-65-2°C. Lower flow rates allowed higher temperatures to be reached at a shorter distance due to longer time of fluid staying at higher temperature (see Fig. S1.2). Higher flow rates ensured faster heating along the first 40 cm of the cell, with a ramp of around 4.6 °C.min$^{-1}$ for the highest flow rate and 0.4 °C.min$^{-1}$ for the lowest one (see Fig S1.3).

The emulsions were passed 4 times through the cell, and samples were taken for drop size measurement after each pass. Figure 3B shows the mean volume diameter of the drops after each cycle and the macroscopic emulsions appearance after the second cycle, while Fig. 3C shows the drop size distributions by volume, after 4 passes at the different flow rates. A similar type of bimodal distribution is present after the first cycle, and the different conditions differ in the efficiency with which they convert the large size droplets to the small ones, though they also affect the final diameter size of the small peak. At 0.1 mL.min$^{-1}$, the coarse emulsion transformed into a translucent nanoemulsion after the first cycle and became transparent after the second cycle (see histograms in Fig. S2.1). The mean drop diameter decreased from 4.8 μm to 151 nm in the first cycle with a bimodal distribution, where 11 vol% of the drops peaked around 1.0 ± 0.6 μm and 89 vol.% peaked at 19.3 ± 8.7 nm (here the "±" sign stands for the peak half-width at half-height). The mean volume size of the micelles in the absence of Precirol was 9.3 nm and their peak was at 8.7 ± 2.6 nm, showing that a substantial fraction of the oil was transformed into nanodroplets, coated by the surfactant in the first pass. After the second pass, the mean size decreased to 69 nm, thus reducing the second peak area to 4 vol% and 2 vol% after the third one.

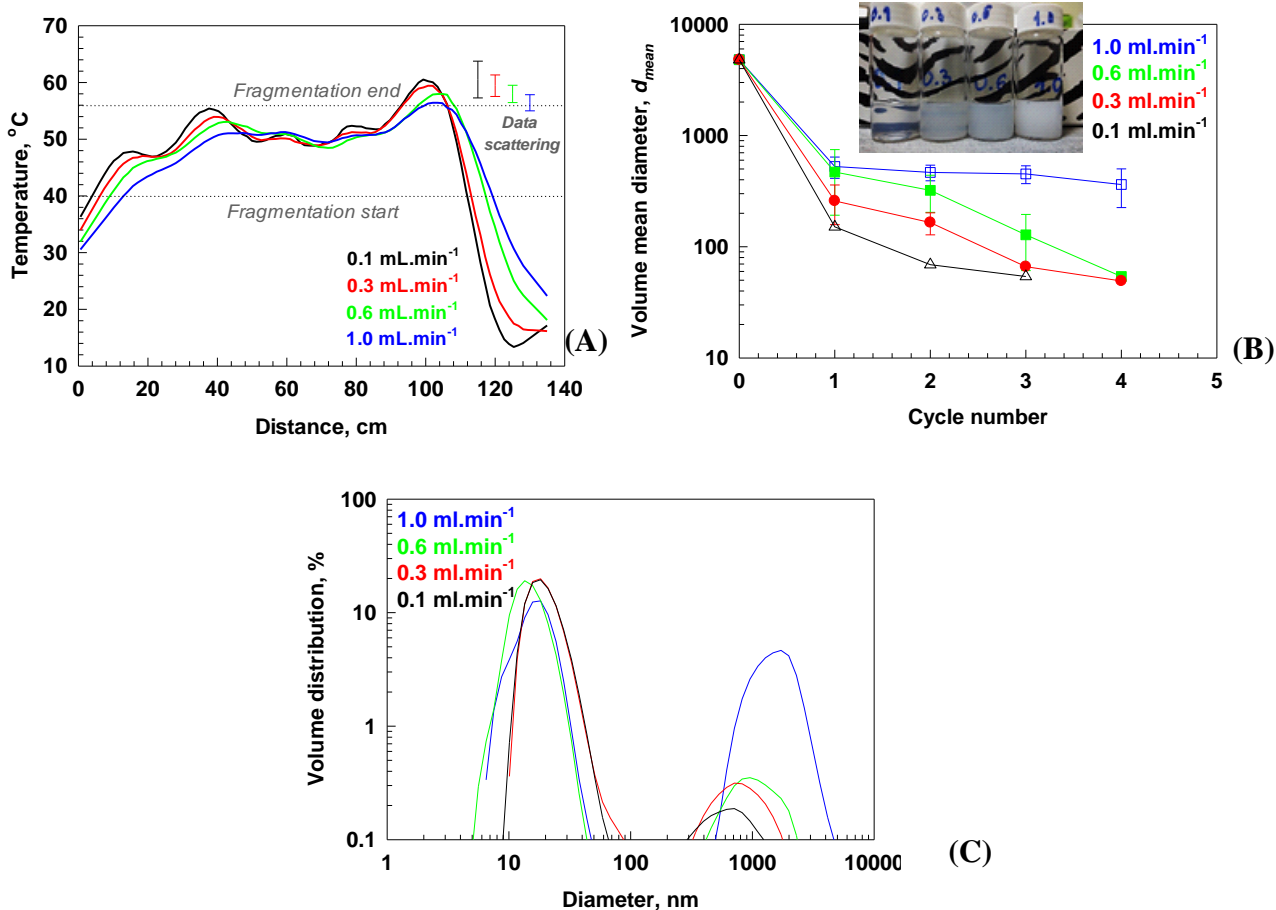

**Figure 3.** **(A)** Temperature profiles in the reactor as a function of the distance in the channel, for different flow rates, as color-coded in the figure. The dotted line illustrates the temperature at which the fragmentation of the particles started via swelling, and the final temperature at which the fragmentation was completed; the vertical error bars in the top-right corner show the maximum temperature fluctuation near the highest temperature point on each curve. **(B)** Mean drop size by volume, as a function of the number of cycles the emulsion passed through the reactor. Error bars represent averaged data from 2 to 5 independent experiments. The inset picture illustrates the emulsion turbidity after the second cycle upon increasing the flow rate from left to right. **(C)** Histograms by volume for emulsion drops after 4 cycles (3 cycles for the 0.1 ml.min$^{-1}$ rate). The studied system is 1 wt% Precirol in 1.5 wt% Brij S20 + 0.5 wt% Brij 30 surfactant solution.

The higher flow rates led to higher nanoemulsion production rates; however, the mean drop size was also larger due to lower efficiency of bursting of the large droplets, as seen by the greater area of the second peak in Fig. 3C. Fragmentation at 1.0 ml.min$^{-1}$ gave the worst result in these tests with 29 ± 9 vol% of the drops remaining in the micrometer range after the first pass, and 21 ± 7 vol% after the fourth pass. The other flow rates gave intermediate results, with 26 and 18 vol% for the 0.6 and 0.3 ml.min$^{-1}$ after the first pass (see Fig S2.1 for more data on volume distributions).

To check whether the slower heating in the emulsion would lead to better fragmentation, we kept the flow rate at 1.0 mL.min$^{-1}$ but varied the temperature gradients as illustrated in Figure 4A. We assessed four temperature profiles and the emulsion was passed through the flow cell only once. Afterward, the drop sizes were measured to compare the breakage efficiency, as illustrated in Figures 4B. As anticipated, the slower heating, allowing longer sample residence time at the relevant temperature interval for bursting, led to better results with only about 23 vol% of the drops remaining at around 2 µm in diameter). In contrast, the more rapidly heated emulsions led to around 34, 40 and 50 vol% of the large drops remaining. Optical microscopy images after the first cycle are presented in Fig. S2.2.

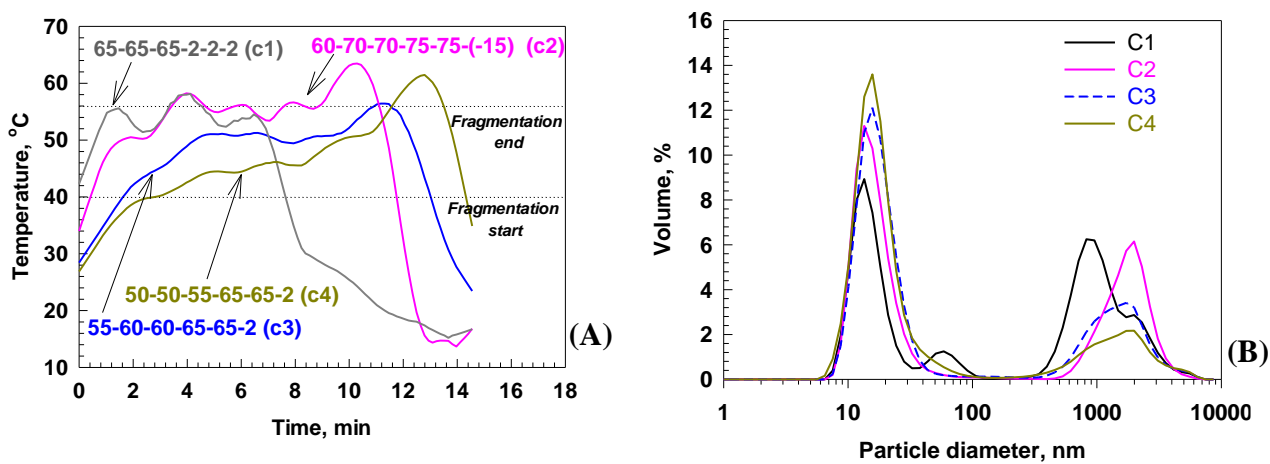

**Figure 4.** **(A)** Temperature profiles at 1 mL.min$^{-1}$, based on the temperatures set in the 6 Peltier elements, as a function of time. The nominal temperature was 26 ± 3°C in all experiments, but due to heat conduction from the cell walls to the connected tubing, a warmer starting temperature was observed for the higher temperature setpoints. The horizontal lines in the figure illustrate the interval, where rapid bursting of the particles occurs, as shown in Fig. 5A. **(B)** Histograms of the drops after 1 pass. The drop diameter before fragmentation was $d_{32}$ = 4.8 µm. The studied system is 1 wt% Precirol in 1.5 wt% Brij S20 + 0.5 wt% Brij 30 surfactant solution.

To better clarify the role of the temperature increase for the fragmentation process, we measured the bulk Precirol endothermic peaks by DSC upon heating and the size of the emulsion particles (by DLS) upon heating after discrete temperature steps. Figure 5A shows the DSC results for the bulk Precirol at a heating rate of 5 °C.min$^{-1}$, closely resembling the rise before the plateau in the temperature profile shown with the blue curve in Figure 4A. Two interconnected peaks were observed: one with maximum at ≈ 48°C and another at ≈ 56°C. Literature data from Refs. [31-32] attribute the two endothermic events for freshly solidified Precirol in DSC experiments to the existence of a mixture of α and β polymorphic forms of the lipid. Indeed, after performing experiments with X-ray scattering (Figure 6a and Figure S3.1a) we observed that two different polymorphs are present upon freezing not only in bulk, but also for emulsified oils. The characteristic lamellar thicknesses measured from SAXS were 7.76 and 4.99 nm for emulsified Precirol. Upon heating, the emulsion sample underwent polymorphic phase transition in 26-38°C temperature region. After that a single broad peak with maxima at 0.129 Å$^{-1}$ (≈ 4.87 nm) was observed until melting. Note that the end of the polymorphic transition denotes the beginning of the fragmentation process as observed under the microscope. Therefore, as previously observed for triglycerides [25,26], in the case of Precirol, the polymorphic phase transitions caused formation of nanocracks, which allowed aqueous phase penetration and particle disintegration.

The complementary DLS measurements (Fig. 5B) allowed us to track the change in the particle size in the same temperature range and with similar heating rates as in Fig 3A. Upon heating from 40 to 45-46°C, the size of the particles slightly increased, indicating a swelling of the lipid particles caused by the penetration of the aqueous phase into the particle cracks, as seen for microparticles in Ref. [26]. This swelling was more pronounced for the slower heating protocols. In the temperature interval, from 46 to 50 °C, the particle size remained nearly constant without changes in the particles' count rate, indicating a slow process that occurred beyond the timescale of this experiment or a dynamic balance between the particle swelling and the fragmentation and diffusion of small nanoparticles from the surface of the particles. Further temperature increase from 50 to 56 °C caused the mean size to decrease sharply, thus reflecting the process of particle fragmentation (bursting) before complete melting above 56 °C.

We calculated the emulsions' residence times in the relevant particle-swelling temperature intervals assuming plug flow and using the original non-smoothed data, which allowed us pinpoint more precisely the transition regions (fragmentation start and end) that we could otherwise omit or shift. To clarify which temperature interval is more important for the particle fragmentation, we plotted the mean drop size from all experiments with Precirol vs. residence time in the intervals: 40-50 °C, where the swelling occurred without fragmentation (or with limited fragmentation); 50-56 °C, where the intensive bursting occurred together with some melting; and the total interval of 40-56 °C, which is shown in detail in S.I. Section S4. Figure 5C shows the data plotted for the interval with the best correlation observed: 40-50 °C, where the particle swelling occurred prior to intensive bursting. To illustrate the role of this interval even better, we replot the results from Figures 3 and 4 in Figure 5C using the color coding from the respective figures. The different cycles are shown as different residence times in the same color, with the residence times being discretely proportional to the number of passes. As observed, the single passing and the multi-cycling with longer residence time in the temperature range of 40-50°C led to the same breakage results into nanoemulsions. This result means that the stage of the particles swelling plays a critical role for the subsequent intensive bursting at higher temperatures.

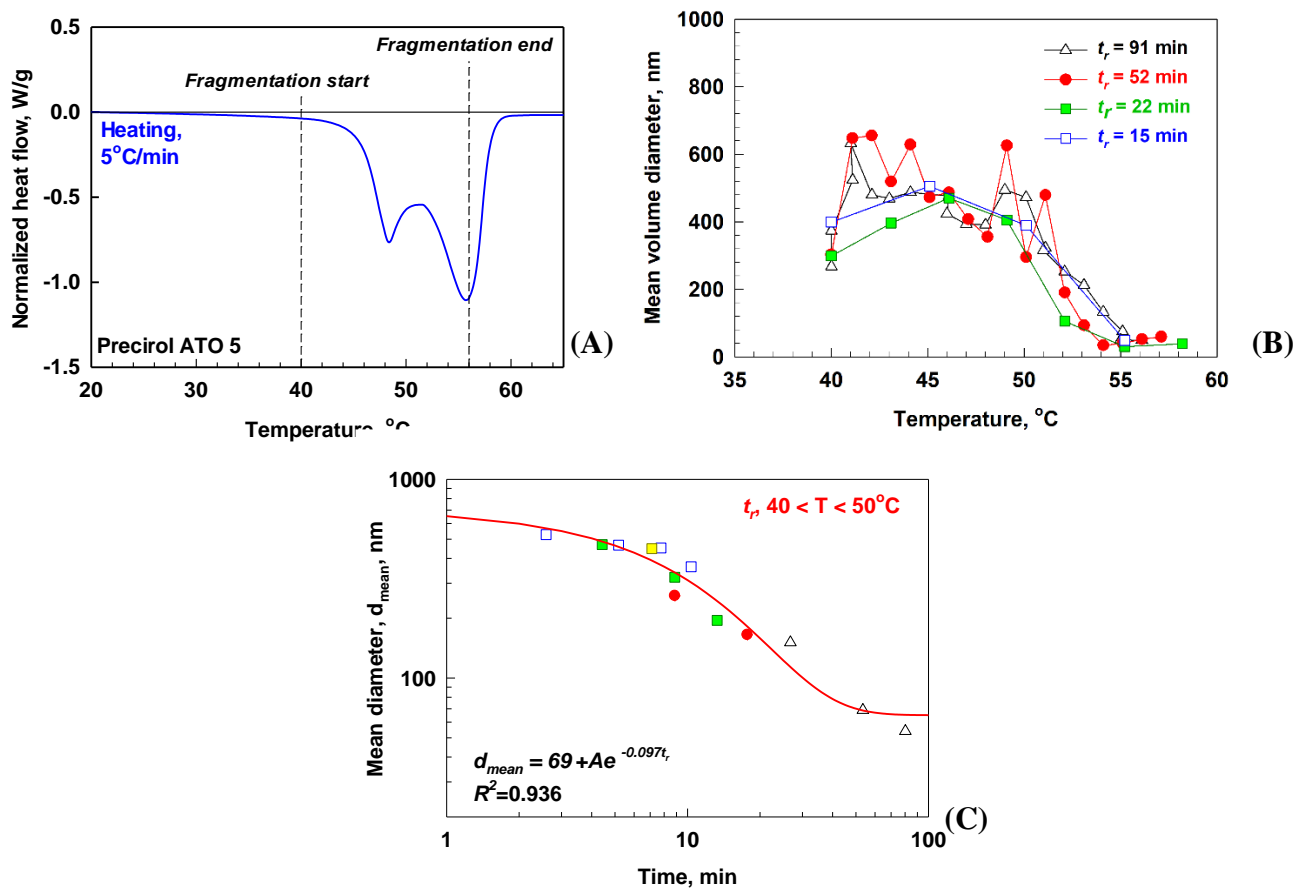

**Figure 5.** (A) Heat flow for bulk Precirol ATO5, as a function of temperature, measured via DSC upon heating (exo up). (B) Mean volume diameter as a function of temperature, for the bursting Precirol ATO5 particles, as measured via DLS, all starting at 40 °C, proceeding at different heating protocols: 1°C temperature step, 5 min measurement time at the given temperature (black triangles); 1°C, 3 min (red circles); 3°C, 3 min (green squares) and 5°C, 3 min (blue squares). The total time for the experiment is given in the legend as $t_r$. The random up-and-down oscillations of the particle size results from the opposite effects of particle swelling and the stochastic particle fragmentation. (C) Mean volume diameter, as a function of the residence time in the polymorphic transition interval for all tested samples. The different times for a given color correspond to different number of passes through the reactor, keeping the color coding from Figs. 3 and 4. Data obtained for 1 wt% Precirol ATO5 dispersed in 1.5 wt% Brij S20 + 0.5 wt% Brij 30 surfactant solution is shown in (B,C).

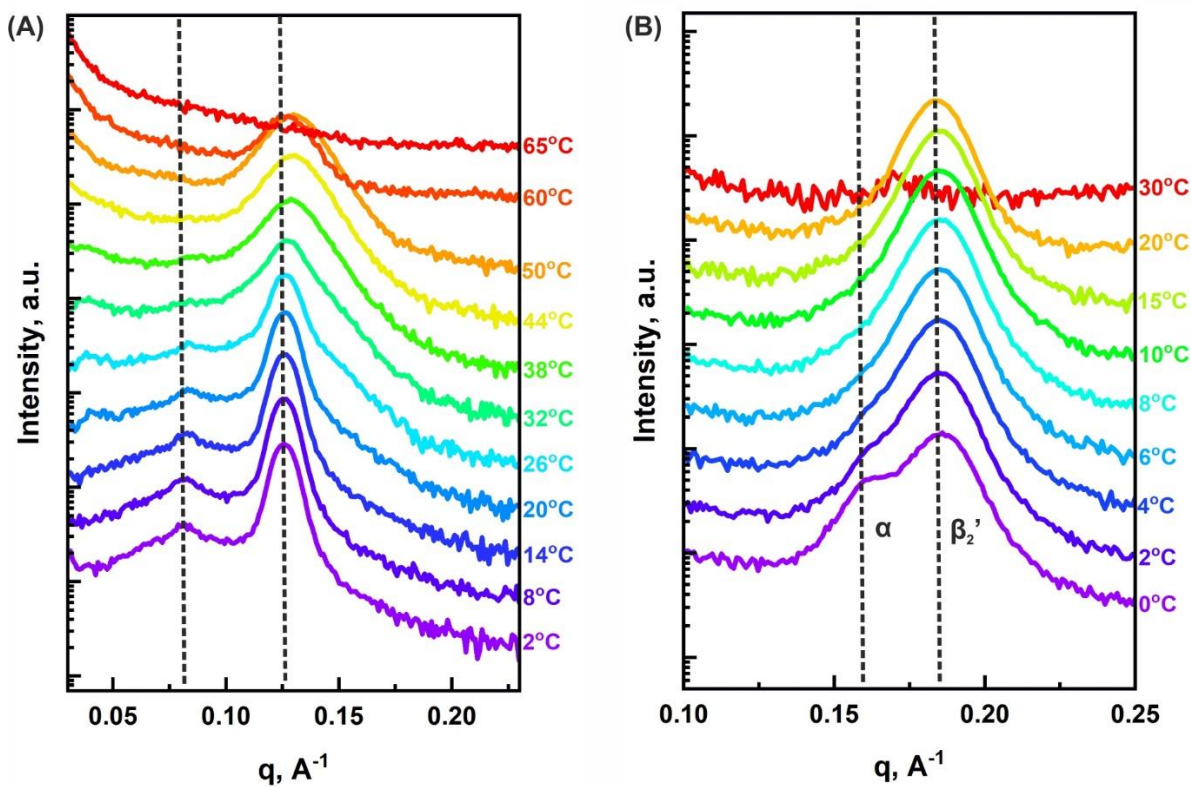

**Figure 6.** SAXS spectra obtained during heating of **(A)** 5 wt% Precirol, dispersed in 4.5 wt% Brij S20 + 1.5 wt% Brij 30 solution and **(B)** 5 wt% coconut oil, dispersed in 4.5 wt% Tween 20 + 1.5 wt% $C_{18:1}MG$ solution. Different colors denote different temperatures, as indicated on the graphs. The samples have been cooled with 30°C/min rate down to 2°C and 0°C respectively and heated with 1.5 and 0.5°C/min rates until melting of the dispersed entities. **(A)** Two distinct peaks are observed with maxima of the scattering vector q ≈ 0.081 Å$^{-1}$ and ≈ 0.126 Å$^{-1}$. Upon heating, a polymorphic transition takes place and the two peaks merge into a single broad peak with maximum ≈ 0.129 Å$^{-1}$. **(B)** Two distinct peaks are observed upon freezing of the CNO emulsion as well. Their maxima are at q ≈ 0.164 and 0.185 Å$^{-1}$. An α to $β_2$' polymorphic transition takes place between ca. 2°C and 10°C. Note that the end of the polymorphic transition marks the beginning of the nanofragmentation process in both samples as observed under optical microscope.

As the swelling occurred while the lipid was still solid, we decided to check if we could enhance surfactant penetration into the particles. With this aim in view, we added 100 mM NaCl in the studied emulsion to increase the osmotic pressure of the continuous phase. Note that 100 mM NaCl is below the isotonic solution concentration (≈ 154 mM), often used in pharmaceutical and medical applications. Then, we ran this emulsion through the reactor using the best temperature profile, i.e. 50-50-55-65-65-2 at 1.0 mL.min$^{-1}$. In Figure 7A we compare the resulting drop size distribution by volume against the respective sample without NaCl added, ran at ×10 lower flow rate, viz. at 0.1 mL.min$^{-1}$ (whereas the histogram at 1.0 mL.min$^{-1}$ without salt was shown in Fig.3C). The histograms look very similar, thus confirming the faster fragmentation in the emulsion with added 100 mM NaCl. The mean drop size in the final emulsions, obtained from the reactor was around 170 nm in the presence of 100 mM NaCl at 1.0 mL.min$^{-1}$ and resembled the drop size distributions obtained with HPH at 500 and 750 bars, where the mean drop sizes after two passes at each pressure were 251 nm and 127 nm, respectively.

We tested also more concentrated emulsions containing 5 wt% Precirol to check if the fragmentation efficiency holds at higher oil fractions. Figure 7B shows the volume distribution after a single reactor pass, compared to the same emulsion at 1 wt% Precirol. The fragmentation in 5 wt % emulsion, which contained 4.5 wt% Brij S20 and 1.5 wt% Brij 30, was highly efficient and the mean drop size was comparable to that in 1 wt% emulsion and to that obtained with HPH at 500 bars (Fig. 7A). We did not study higher oil fractions to avoid reactor clogging, but conceptually it should be possible when more efficient pumps are used.

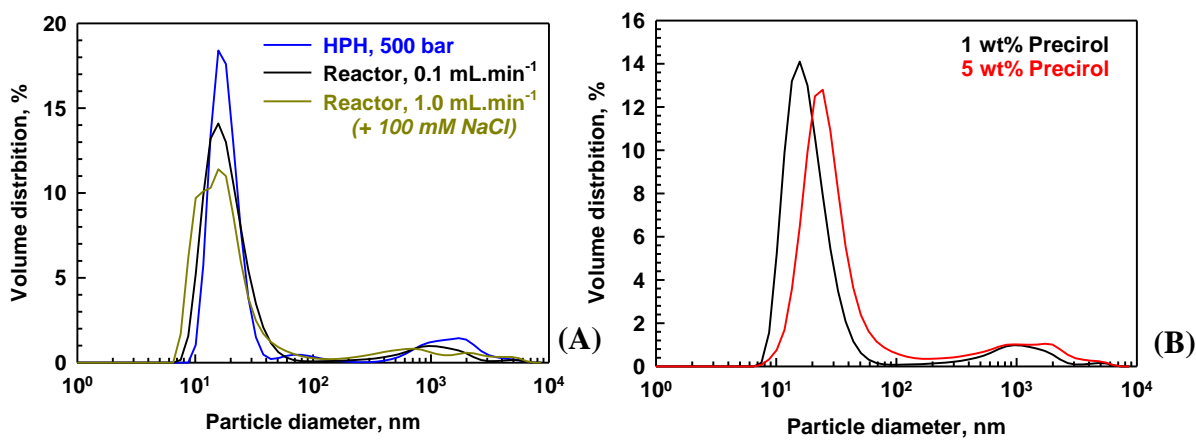

**Figure 7.** **(A)** Histograms by volume for 1 wt% Precirol ATO 5 drops dispersed in 1.5 wt% Brij S20 + 0.5 wt% Brij 30 + 100 mM NaCl after 1 pass through the reactor at *T*-sequence 55-60-60-65-65-2 °C and 0.1 mL.min$^{-1}$, and at 50-55-55-65-65-2 °C at 1.0 mL.min$^{-1}$. The results are compared against HPH at 500 bars. **(B)** Histograms by volume for emulsions with different fractions of Precirol at *T*-sequence 55-60-60-65-65-2 °C and 0.1 mL.min$^{-1}$.

Next, we tested the reactor for the fragmentation of 1 wt% emulsion of the natural coconut oil in the presence of the surfactants 1.5 wt% Tween 20 and 0.5 wt% monoolein glyceride. The coconut oil consists of a wide range of triglycerides and its α-polymorphic freezing peak is at ≈ 4°C [26]. Therefore, we set temperature profiles that lead to freezing of the emulsion drops before heating them in the second half of the cycle, Figure 8A. We used two temperature profiles: one that cooled the emulsion down to 1.3 °C, followed by slow heating at ≈ 0.03 °C/min; and second one in which -3.3°C were reached as a minimum temperature without water freezing by adding 2

M NaCl in the aqueous phase. We passed the emulsion once through the reactor and measured the drop size distribution as shown in Fig. 8B. The mean volume diameter was 249 nm, slightly larger than the one obtained with HPH at 500 bars (151 nm) and significantly smaller than the HPH result at 300 bars (636 nm).

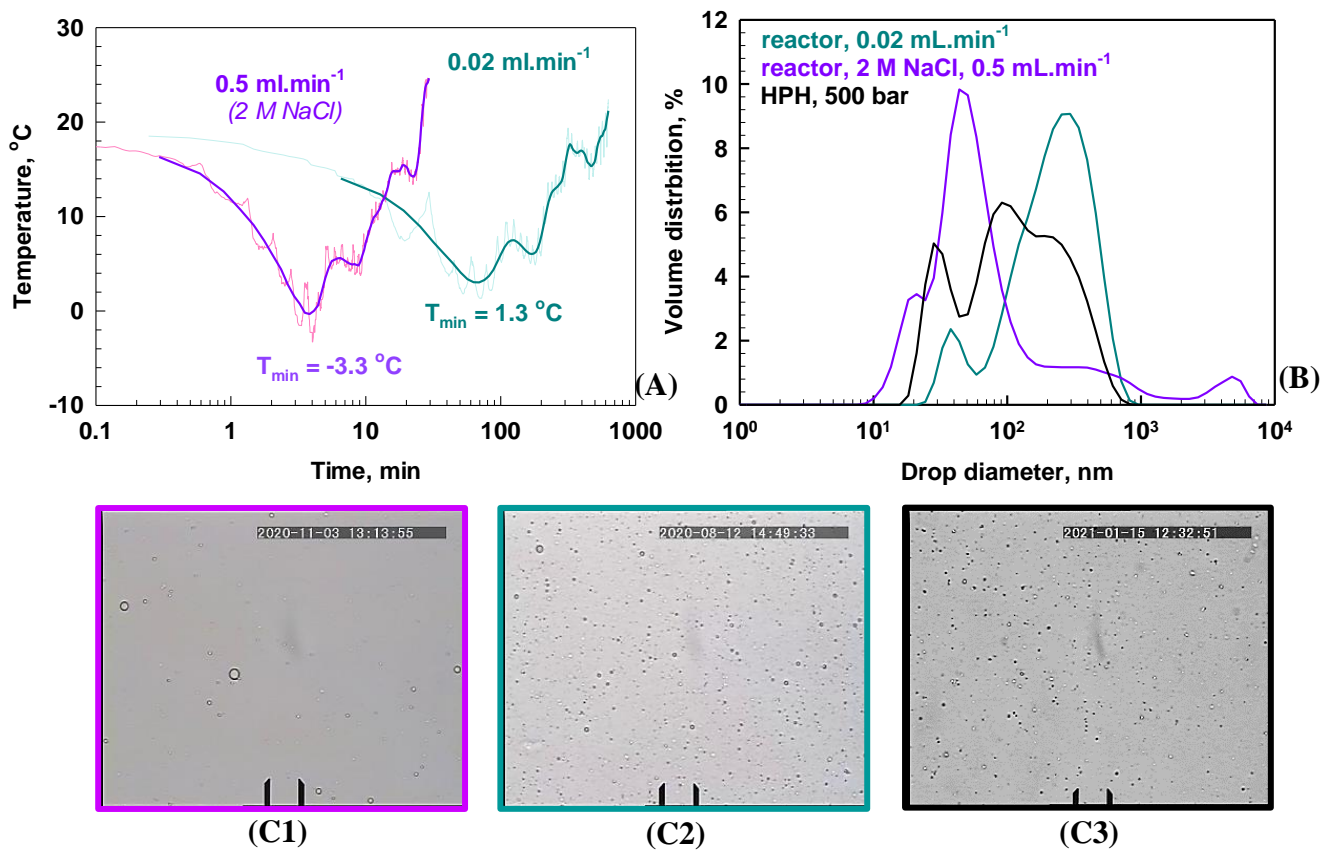

**Figure 8.** **(A)** Temperature profiles, as a function of time for a cooling protocol (-5)-(-5)-10-15-15-20 °C at 0.02 mL.min$^{-1}$, and (-15)-(-15)-7-15-15-30 °C at 0.50 mL.min$^{-1}$. **(B)** Histograms by volume for emulsions prepared via HPH or via reactor, set as in (A). **(C)** Optical images from the emulsions, presented in (B). Scale bars are 20 μm. 1 wt% CNO drops dispersed in 1.5 wt% Tween 20 + 0.5 wt% Monoolein have been used.

It should be noted here that adding 100 mM NaCl to a 0.5 ml.min$^{-1}$ profile at a temperature protocol (-5)-(-5)-10-15-15-20 °C, i.e. as in the cyan curve in Fig 8A, was not sufficient to break the droplets into smaller ones, and mean diameter of 686 ± 245 nm was obtained. The latter result implies that lower temperature is necessary to overcome the severe supercooling of the coconut oil before complete freezing [26].

To further explore the reactor's capabilities, we studied emulsions prepared from the natural Palm kernel oil, coconut oil (CNO) and the pure trimyristin. We varied the surfactant compositions and the flow rates to clarify the role of these variables and to optimize the particle fragmentation.

In Figure 9 we summarize the mean particle diameters, plotted as a function of the residence time of the emulsion upon heating in the reactor. The temperature intervals, used for determining the swelling interval before rapid bursting and the residence times for the different temperature profiles were calculated as for the Precirol data, illustrated in Fig. 5 and S.I. Figure S4.1. Depending on the oil composition, we observed two clear trendlines – oils with a more complex composition, such as the coconut and palm kernel oil, required about 10 times longer residence times upon heating to achieve the same breakage when compared to the simpler lipids, e.g. the trimyristin and Precirol. Here we should note that the dataset for Precirol was further optimized to determine the swelling interval alone, via empirical correlations shown in S.I. Figure S4.2. Even though, the complex oils might benefit from such an optimization, a more prominent effect was addressed instead. Namely, the osmotic pressure of the aqueous phase. When it was increased by adding NaCl, the fragmentation efficiency of the Precirol and coconut oil increased, thus significantly shortening the needed residence time even further. However, the coconut oil always required significant supercooling below 0°C, and longer residence times for efficient particle fragmentation. We note that in this case the nanofragmentation was also related to the polymorphic transitions proceeding in the samples, as demonstrated in SAXS/WAXS measurements, see Figure 6b and Figure S3.1b.

We note also that the transparent nanoemulsions, such as the Precirol emulsion obtained at low flow rates and the coconut oil emulsion with the smallest drops are stable within several months without detectable changes in the droplet/particles size (the stability of the coconut emulsions with 2 M NaCl was not investigated on the long term). Similar effects were studied also in Refs. [25-26] with batch emulsions.

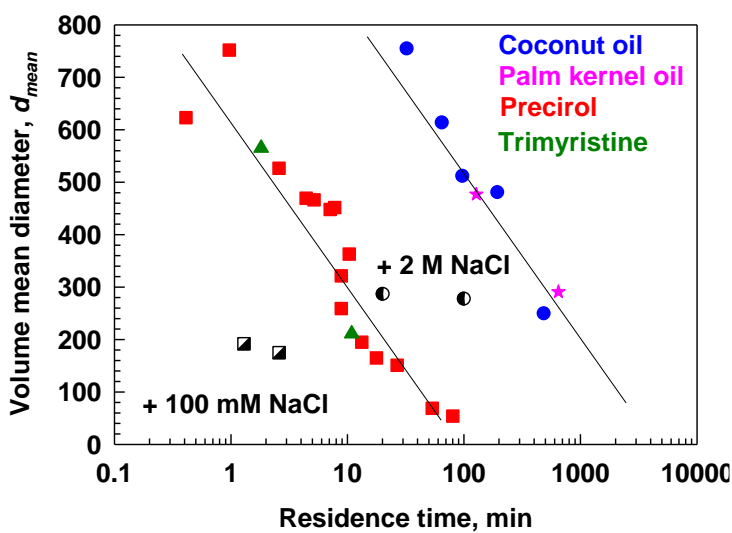

**Figure 9.** Mean particle diameter by volume, as a function of the particle residence time for emulsions obtained at different temperature profiles, flow rates and oils: coconut oil (circles), palm

kernel oil (stars), Precirol (squares) and trimyristin (triangles). The semi-filled squares are for Precirol in the presence of 100 mM NaCl and the semi-filled circles are for coconut oil at 2 M of NaCl. Temperature profiles and temperature intervals used for the evaluation are shown in the Supplementary Material, Section S4. Lines are guides to the eye.

## 4. Conclusions

We developed a pilot scale reactor for continuous flow fragmentation of lipids into nanoemulsions and nanosuspensions via polymorphic phase transitions in the lipid phase. The reactor settings were optimized to achieve efficient fragmentation with four oils, by tuning the temperature protocols, the injecting flow rate, and the surfactants and electrolytes used. Shorter residence times, 10-100 min, were sufficient for some of the oils (Precirol and trimyristin), whereas 100-1000 min were needed for the natural triglycerides, depending on the presence of electrolytes. The fragmentation process was notably accelerated by increasing the solution osmotic pressure which could be achieved by adding NaCl or other solutes.

The current reactor allows one to produce nanoemulsions continuously, based on the optimized temperature profiles. It can be scaled up to an industrial level using the same temperature profiles and time scales, as in the current equipment. The productivity could be enhanced significantly by increasing the diameter of the flow channel and by using multiple flow channels in parallel. The specific optimization is strongly dependent on the emulsion swelling interval prior to bursting and melting, and the desired final size of the particles.

Summarizing, the conclusions drawn above allow one to construct an optimized, highly efficient reactor by selecting the most appropriate temperature range and residence time of the droplets, and the electrolyte concentration.

## Contributions

**Ivan Lesov:** Methodology; Investigation; Validation; Formal analysis; Visualization; Writing – original draft; **D. Glushkova**: Investigation; Visualization; **D. Cholakova:** Methodology; Investigation (SAXS/WAXS); **M.Georgiev:** Methodology; **S.Tcholakova**: Conceptualization; Methodology; Supervision; Funding acquisition; Project administration; **S. Smoukov**: Conceptualization; Funding acquisition; Project administration; Writing – review and editing; **N. Denkov:** Conceptualization; Methodology; Supervision; Writing – review and editing## Acknowledgments

The authors express their gratitude to Mr. Bozhidar Ivanov for the assistance in the reactor design and construction.

**Funding**

The authors gratefully acknowledge the financial support of the following grant to SKS: Proof-of-concept grant CoolNanoDrop (# 841827) by the European Research Council (ERC).

This work is partially supported by the Operational Program "Science and Education for Smart Growth", Bulgaria, grant number BG05M2OP001-1.002-0012.

**Declaration of Competing Interests**

The authors declare that they have no known competing financial interests or personal relationships that could have influenced the work reported in this paper.

**Supplementary data is available for the current manuscript.**